\documentclass[11pt,a4paper]{article}
\usepackage{amssymb,epsfig}
\usepackage{times,cite}
\newcommand{\be}{\begin{equation}}
\newcommand{\ee}{\end{equation}}
\newcommand{\bea}{\begin{eqnarray}}
\newcommand{\eea}{\end{eqnarray}}
\newcommand{\pa}{\partial}

\newcommand{\A}{\mathbb{A}}
\newcommand{\C}{\mathbb{C}}

\newcommand{\N}{\mathbb{N}}

\newcommand{\ci}{\bullet}

\newcommand{\bfg}{\begin{figure}}
\newcommand{\efg}{\end{figure}}
\newcommand{\bc}{\begin{center}}
\newcommand{\ec}{\end{center}}

\title{From nonassociativity to solutions of the KP hierarchy\thanks{\copyright 2006 
by A. Dimakis and F. M\"uller-Hoissen} \, \footnote{Presented 
at the International Colloquium ``Integrable Systems and Quantum Symmetries'', 
Prague, 15-17 June 2006.}}

\author{Aristophanes Dimakis \\
 Department of Financial and Management Engineering, \\
 University of the Aegean, 31 Fostini Str., GR-82100 Chios, Greece \\
 dimakis@aegean.gr
          \and
 Folkert M\"uller-Hoissen \\ Max-Planck-Institute for Dynamics and Self-Organization \\
 Bunsenstrasse 10, D-37073 G\"ottingen, Germany \\
 folkert.mueller-hoissen@ds.mpg.de }

\date{}

\begin{document}

\maketitle
\begin{abstract}
A recently observed relation between `weakly nonassociative' algebras $\A$ (for which 
the associator $(\A,\A^2,\A)$ vanishes) and the KP hierarchy (with dependent variable 
in the middle nucleus $\A'$ of $\A$) is recalled. For any such algebra there is 
a nonassociative hierarchy of ODEs, the solutions of which determine solutions of the KP hierarchy. 
In a special case, and with $\A'$ a matrix algebra, this becomes a matrix Riccati 
hierarchy which is easily solved. The matrix solution then leads to solutions of 
the scalar KP hierarchy. We discuss some classes of solutions obtained in this way.
\end{abstract}

\section{Introduction} 
Let us call an algebra $\A$ (over a commutative ring) \emph{weakly nonassociative (WNA)}  
\cite{DMH} if 
\be
    (a,b \, c,d) = 0  \qquad \quad  \forall a,b,c,d \in \A  \, ,  \label{WNA}
\ee
where $(a,b,c) := (a \, b) \, c - a \, (b \, c)$ is the associator in $\A$. 
The \emph{middle nucleus} of $\A$, 
\be
    \A' := \{ b \in \A  \; | \; (a,b,c) =0 \; \; \forall a,c \in \A \} \, , 
\ee 
is an \emph{associative} subalgebra and a two-sided ideal. Let us fix 
$f \in \A \setminus \A'$ and define $a \circ_1 b := a \, b$ and\footnote{As a consequence 
of (\ref{WNA}), these products only depend on the equivalence class $[f]$ of $f$ in $\A/\A'$.} 
\be
        a \circ_{n+1} b := a \, (f \circ_n b) - (a \, f) \circ_n b  
        \qquad \quad  n=1,2, \ldots \; .
\ee
The subalgebra $\A(f)$, generated in $\A$ by $f$, is called \emph{$\delta$-compatible} if, 
for all $n \in \N$, 
\be
        \delta_n(f) := f \circ_n f 
\ee
extends to a \emph{derivation} of $\A(f)$. 
\vskip.1cm

\noindent
{\bf Theorem 1.} \cite{DMH} Let $\A(f)$ be $\delta$-compatible. The derivations $\delta_n$ 
then commute on $\A(f)$ and satisfy identities which are in correspondence 
(via $\delta_n \mapsto \pa_{t_n}$) with the equations of the potential KP hierarchy 
(with dependent variable in $\A'$). 
\hfill $\square$
\vskip.1cm

\noindent
{\bf Example.} The first three derivations are determined by 
\bea
  \delta_1(f) = f^2 , \; \, 
  \delta_2(f) = f \, f^2 - f^2 \, f , \; \, 
  \delta_3(f) = f \, (f \, f^2) - f \, f^2 \, f - f^2 f^2 + (f^2 f) \, f 
\eea
and the derivation rule $\delta_n(ab) = \delta_n(a) \, b + a \, \delta_n(b)$. 
They satisfy the identity
\be
   \delta_1 \left( 4 \, \delta_3(f) - \delta_1^3(f) + 6 \, (\delta_1(f))^2 \right)
   - 3 \, \delta_2^2(f) + 6 \, [\delta_2(f) , \delta_1(f) ] \equiv 0  \, ,
\ee
which via $\delta_n \mapsto \pa_{t_n}$ becomes the \emph{potential KP equation} 
(for $-f$).\footnote{Our conventions correspond to `KPII'. In section~3 we 
also consider `KPI' which is obtained from KPII via $t_{2n} \mapsto \imath \, t_{2n}$. 
For water waves, KPI applies to the case where surface tension dominates over gravity.}
\hfill $\square$

The result formulated in theorem~1 provides us with a way to obtain 
solutions of the KP hierarchy by solving \emph{ordinary} differential 
equations (ODEs). 
\vskip.1cm

\noindent
{\bf Theorem 2.} \cite{DMH} Let $\A$ be any WNA algebra over the ring of functions 
of independent variables $t_1,t_2, \ldots$. If $f \in \A$ solves the 
hierarchy\footnote{The flows given by (\ref{na_odes}) indeed commute \cite{DMH}. 
$f$ has to be differentiable, of course, which requires 
a corresponding (e.g. Banach space) structure on $\A$. 
If $f$ solves (\ref{na_odes}), the algebra $\A(f)$ generated by $f$ in $\A$ 
over the subring of constants is $\delta$-compatible \cite{DMH}.}
\be
   \pa_{t_n}(f) = f \circ_n f  \qquad \qquad   n=1,2, \ldots   \label{na_odes}
\ee
of ODEs, then $-\pa_{t_1}(f)$ lies in $\A'$ and solves the \emph{KP hierarchy} 
(with dependent variable in $\A'$). If there is a \emph{constant} 
$\nu \in \A \setminus \A'$ with $[\nu] = [f] \in \A/\A'$, then 
\be
    \phi := \nu - f \in \A' 
\ee
solves the \emph{potential KP hierarchy}. 
\hfill $\square$

In order to apply this result, we need to know more about WNA algebras. 
For our purposes, it is sufficient to recall from \cite{DMH} that 
any WNA algebra with $\mathrm{dim}(\A/\A')=1$ is isomorphic to one 
determined by the following data: \\
(1) an associative algebra $\mathcal{A}$ (e.g. any matrix algebra) \\
(2) a fixed element $g \in \mathcal{A}$ \\
(3) linear maps $L, R \, : \, \mathcal{A} \rightarrow \mathcal{A}$ such that 
\be  
    [L,R] =0 \, , \qquad 
    L(a \, b) = L(a) \, b \, , \qquad 
    R(a \, b) = a \, R(b) \; .  \label{LR-rels}
\ee
Augmenting $\mathcal{A}$ with an element $\nu$ such that
$\nu \, \nu = g$, $\nu \, a = L(a)$, $a \, \nu = R(a)$, leads to 
a WNA algebra $\A$ with $\A' = \mathcal{A}$.

\section{A class of WNA algebras and corresponding KP solutions}
Let $\mathcal{A} = \mathcal{M}(M,N)$ be the algebra of complex $M \times N$ matrices 
with the product
\be
     A \ci A' := A \, K \, A'  \, ,
\ee
where $K$ is a fixed $N \times M$ matrix. We define linear maps $L,R$ 
as multiplication (from left, respectively right) by a constant $M \times M$ matrix $L$, 
respectively a constant $N \times N$ matrix $R$. Then (\ref{LR-rels}) holds. 
Furthermore, we introduce the new product
\be
     A \circ_1 A' := (A R) \ci A' - A \ci (L A') = A \, (RK -KL) \, A' 
\ee
and augment $(\mathcal{A}, \circ_1)$ by an element $\nu$ such that\footnote{Here we 
set $g=0$, which is a restriction of the possibilities in the case under consideration.}
\be
     \nu \circ_1 \nu = 0 \, , \quad
     \nu \circ_1 A = L \, A \, , \quad
     a \circ_1 \nu = - A \, R \, ,    \label{nu_augm}
\ee
to obtain a \emph{WNA algebra} $(\A,\circ_1)$ with $\A' = \mathcal{A}$.
The reason for resolving the WNA product $\circ_1$ in terms of $\ci$ is the 
drastic simplification in 
\be  
    A \circ_n A' = A \, ( R^n K - K L^n ) \, A' \, ,
\ee
which turns the hierarchy of ODEs (\ref{na_odes}) into the special matrix Riccati equations
\be
  \pa_{t_n}(\phi) = L^n \phi - \phi \, R^n + \phi \, (K L^n - R^n K) \, \phi 
  \qquad n=1,2,\ldots  \; .
\ee
They are easily solved:
\be  
   \phi = -(I_M + e^{\xi(L)} \, C \, e^{-\xi(R)} \, K)^{-1} \, e^{\xi(L)} \, C \, e^{-\xi(R)}
        = -e^{\xi(L)} \, (I_M + B)^{-1} \, C \, e^{-\xi(R)} 
\ee
with a constant matrix $C \in \mathcal{M}(M,N)$, the $M \times M$ unit matrix $I_M$, and 
\be
    B := C \, e^{-\xi(R)} \, K \, e^{\xi(L)} \, , \quad
       \xi(L) := \sum_{n \geq 1} t_n \, L^n  \; .
\ee
According to theorem~2, $\phi$ solves the matrix potential 
KP hierarchy in $(\mathcal{A}, \circ_1)$, thus
\be  
   \varphi := \phi \, (RK -KL)
\ee
solves the matrix potential KP hierarchy with the ordinary matrix product.\footnote{This 
follows e.g. immediately from a functional representation of the potential KP hierarchy 
\cite{DMH,BK}.} 
If $-C$ is the unit matrix, such a solution appeared in \cite{CS} 
in the context of an operator approach towards solutions of scalar nonlinear equations. 
The basic idea is to associate with the respective nonlinear (soliton) equation an operator 
(e.g. matrix) version, to look for exact solutions of the latter and a homomorphism 
into scalars, which then determines solutions of the scalar nonlinear (soliton) equation 
(see also \cite{Mar}). 
In fact, in the case under consideration such a homomorphism 
is obtained as described below, if $K,L,R$ are such that $\mathrm{rank}(RK-KL)=1$ 
(see also \cite{CS,rank1}). 
The latter condition means that there is a $v \in \C^M$ and a $w \in \C^N$ 
with $R K - K L = w \, v^T$. Then the map from $\mathcal{A}$ to smooth 
functions of $t_1,t_2,\ldots$, defined by
\be
   \Psi(A) := v^T A \, w = \mathrm{tr}(A \, w v^T)
            = \mathrm{tr}( A \, (RK-KL) )  \, ,
\ee
has the homomorphism property\footnote{More generally, one can construct 
homomorphisms into matrix algebras in a similar way \cite{DMH}.}
$\Psi(A \circ_1 A') = \Psi(A) \, \Psi(A')$. 
As a consequence, a solution of the \emph{scalar} KP hierarchy is given by 
\be
     u := \Psi(\phi)_{t_1} = \mathrm{tr}(\varphi) 
        = (\mathrm{log} \, \tau)_{t_1 t_1}  \quad \mbox{with} \quad 
  \tau := \mathrm{det}(I_M + B) \; . 
\ee
Note that $\tau$ is in particular invariant under 
\be
    C \mapsto P \, C \, \tilde{P}^{-1} \, , \quad
    K \mapsto \tilde{P} \, K P^{-1} \, , \quad
    L \mapsto P \, L \, P^{-1} \, , \quad
    R \mapsto \tilde{P} \, R \, \tilde{P}^{-1} \, ,
\ee
with any constant invertible $M \times M$ matrix $P$ and $N \times N$ matrix $\tilde{P}$. 
This can be used to reduce both, $L$ and $R$, to Jordan normal form.

\section{Some solutions of the scalar KP hierarchy} 
Let $L = \mathrm{diag}(p_1,\ldots,p_M)$, $R = \mathrm{diag}(q_1,\ldots,q_N)$, and 
\be
         K_{ij} = (q_i - p_j)^{-1} \, ,    \label{K-sol}
\ee
assuming $q_i \neq p_j$, $i=1,\ldots,N$, $j=1,\ldots,M$. Then 
$\mathrm{rank}(RK-KL)=1$ and 
\be
     B = C \, E \qquad \mbox{where} \quad
    E_{ij} := e^{\xi(p_j)-\xi(q_i)}/(q_i - p_j)  \; .
\ee

If $M=N$ and with \emph{real} $L,R$ and $C = \mathrm{diag}(c_1,\ldots,c_N)$, $u$ becomes 
an \emph{$N$-soliton} solution of the scalar KP hierarchy \cite{Nsol,MZBIM,MS}. 
For example, with $M=N=2$, $p_1=-p_2 = \alpha + \beta$, $q_1=-q_2 = \alpha - \beta$, 
$C = (2 \alpha \beta/\sqrt{\alpha^2-\beta^2}) \,\mathrm{diag}(-1,1)$, we obtain
\be
   \tau = 2 \, e^{4 \alpha \beta \, t_2} \, [ (\alpha/\sqrt{\alpha^2-\beta^2}) \,  
         \cosh\left( 2 \beta \, ( t_1 + (3 \alpha^2 + \beta^2) \, t_3) \right) 
         + \cosh(4 \alpha \beta \, t_2) ]  \label{tau_2solitons}
\ee
for $t_4,t_5,\ldots =0$. We may drop the exponential factor, since it does 
not influence $u$. 
The corresponding solution of the KPII equation, which is 
regular if $\alpha>0$ and $|\alpha|>|\beta|$, is shown in Fig.~1.\footnote{The 
plots in this work were generated with Mathematica \cite{Math}.}  
\bfg[t] 
\bc 
\resizebox{!}{3cm}{
\includegraphics{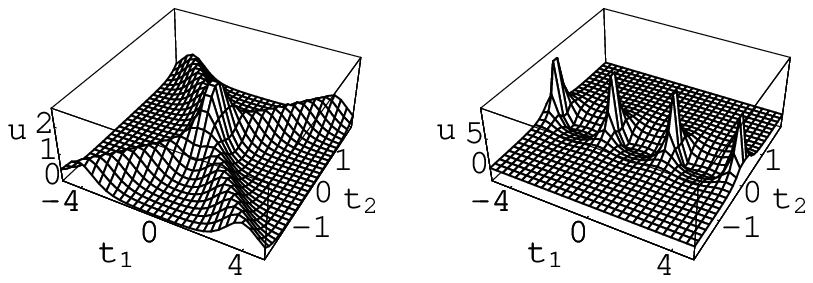}
}
\ec 
\vspace{-10mm}
\caption{A 2-soliton solution of KPII given by (\ref{tau_2solitons}) 
and an associated periodic KPI solution given by (\ref{tau_period}) 
at $t_3=0$, where $\alpha=1.1, \beta=1$.}
\efg   
 For $M=N=2$ and no restriction on $C$, we have
\bea
   \tau &=& 1 + \frac{C_{11} \, e^{\xi(p_1)-\xi(q_1)}}{q_1-p_1} 
              + \frac{C_{12} \, e^{\xi(p_1)-\xi(q_2)}}{q_2-p_1} 
              + \frac{C_{21} \, e^{\xi(p_2)-\xi(q_1)}}{q_1-p_2} 
              + \frac{C_{22} \, e^{\xi(p_2)-\xi(q_2)}}{q_2-p_2} \nonumber \\
        & & - \mathrm{det}(C) \, \frac{(p_2-p_1)(q_2-q_1)}{(q_1-p_1)(q_2-p_1)(q_1-p_2)(q_2-p_2)} \, 
               e^{\xi(p_1)+\xi(p_2)-\xi(q_1)-\xi(q_2)}  \, .
\eea
Let us fix an order: $p_1<p_2<q_1<q_2$. Then $\tau$ is positive for all 
$t_1,t_2,\ldots \in \mathbb{R}$, so that the KP solution $u$ is regular, if 
$C_{ij} > 0$ and $\mathrm{det}(C)<0$. 
Let $1 \leq n \leq 5$ be the number of linearly independent exponential terms in 
this expression. For $n=1$ we have a 1-soliton solution. For $n=2$, this is a 
\emph{Miles resonance} \cite{Mil}, a `Y junction'. 
For $n=3$ we have an ordinary 2-soliton solution and for $n=4$ the type of resonance shown 
in Fig.~2.\footnote{Using instead of $t_3$ any 
other `evolution time' from $t_4,t_5,\ldots$ only means a difference in the velocity.}  
For $n=5$ the behavior of the solution is shown in Fig.~3. 
\bfg[t] 
\bc 
\resizebox{!}{3cm}{
\includegraphics{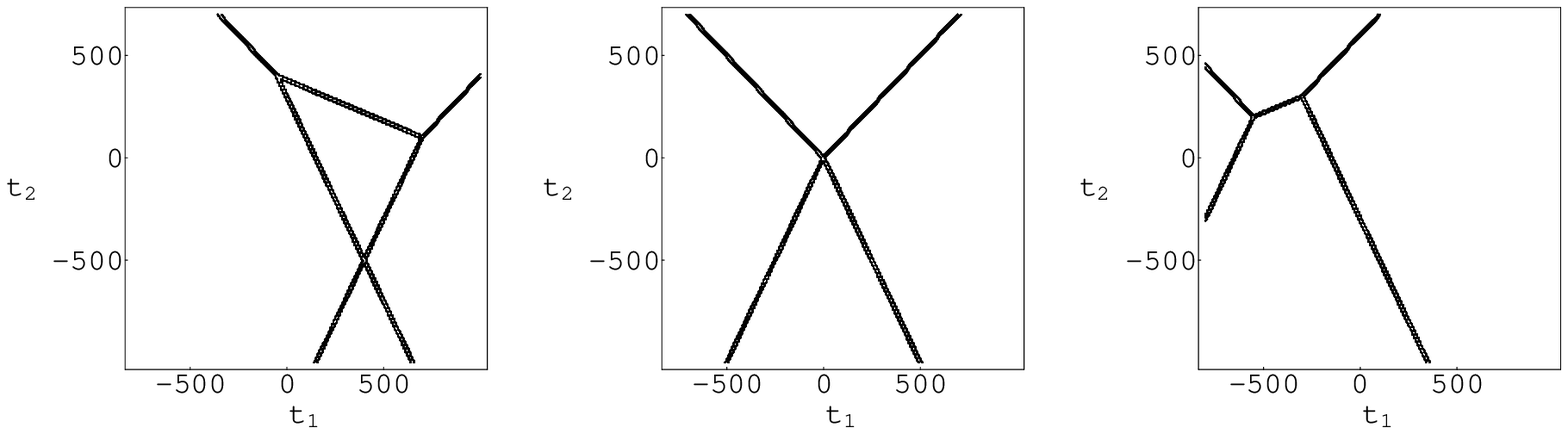}
}
\ec 
\vspace{-10mm}
\caption{Contour plot of a 2-soliton resonance with $p_1=-2,p_2=-\frac{1}{2},q_1=1, q_2=\frac{3}{2}$, $C_{11}=C_{12}=C_{21}=1,C_{22}=0$ at $t_3=-200, 0, 200$, respectively.}
\efg   
\bfg[t] 
\bc 
\resizebox{!}{3cm}{
\includegraphics{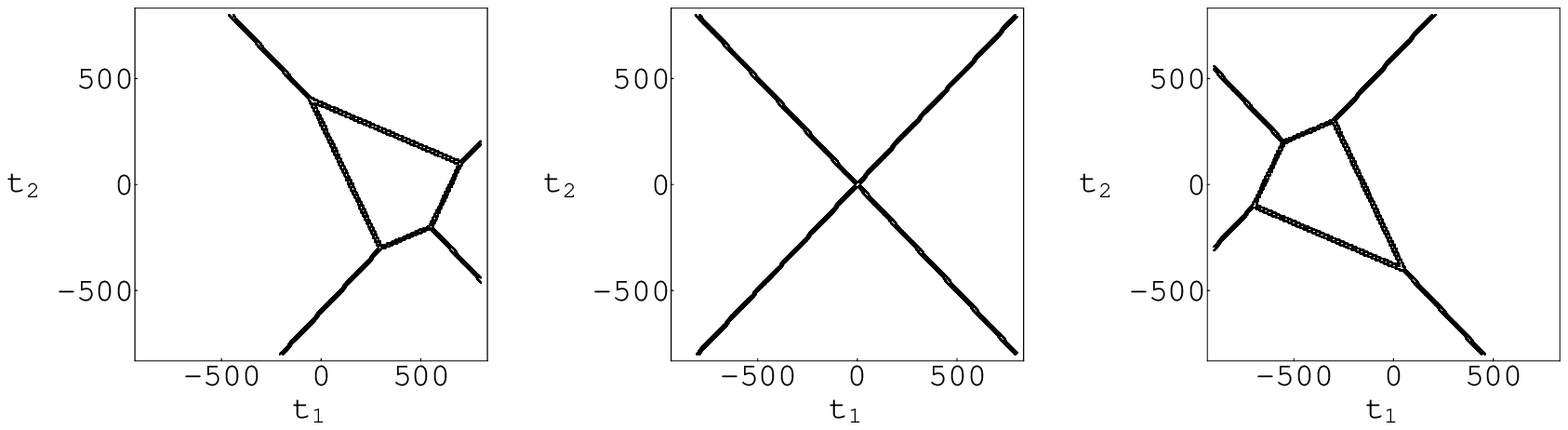}
}
\ec 
\vspace{-10mm}
\caption{Contour plot of a 2-soliton resonance with $p_1=-2,p_2=-\frac{1}{2},q_1=1, q_2=\frac{3}{2}$,  $C_{11}=C_{12}=C_{22}=1,C_{21}=2$ at $t_3=-200, 0, 200$, respectively.}
\efg   
With other values of $M$ and $N$, and real $L,R,C$, one obtains further 
line soliton resonances. For $M=1, N=2$, we have 
$\tau = 1 + ( \alpha \, e^{-\xi(q_1)} + \beta \, e^{-\xi(q_2)} ) \, e^{\xi(p_1)}$, 
where $\alpha := C_{11}/(q_1-p_1)$, $\beta := C_{12}/(q_2-p_1)$. Then  
$u$, which is regular if $\alpha,\beta >0$, is a Miles resonance. 
More involved examples are easily generated 
\cite{K,BC}\footnote{The relation with the expression for $\tau$ functions in 
\cite{BC} is as follows. First we write 
$\tau = \mathrm{det}(I_M + B) = \tilde{\tau} \, \mathrm{det}(e^{\xi(L)})$ with 
$\tilde{\tau} := \mathrm{det}( e^{-\xi(L)} + C \, e^{-\xi(R)} \, K ) 
= \mathrm{det}( \tilde{C} \, \Theta \, \tilde{K}^T )$, where 
$\tilde{C} := (I_M , C)$, $\tilde{K} := (I_M, K)$ and 
$$
   \Theta := \left(\begin{array}{cc} e^{-\xi(L)} & 0_{M \times N} \\
        0_{N \times M} & e^{-\xi(R)} \end{array}\right) \; .
$$
Since $u = (\log(\tau))_{t_1t_1} = (\log(\tilde{\tau}))_{t_1t_1}$, we may replace 
$\tau$ by $\tilde{\tau}$. Multiplication of $\tilde{C}$, and independently $\tilde{K}$, 
from the left by any constant invertible $M \times M$ matrix does not change $u$. 
}.

An example with \emph{complex} $L,R$ is obtained from (\ref{tau_2solitons}) 
with the substitutions $t_2 \mapsto \imath t_2, \beta \mapsto \imath \beta$ 
(after dropping the exponential factor):
\be
   \tau = (\alpha/\sqrt{\alpha^2 + \beta^2}) \,  
         \cos\left( 2 \beta \, (t_1 + (3 \alpha^2 - \beta^2) \, t_3 \right) 
         + \cosh(4 \alpha \beta \, t_2) \; .  \label{tau_period}
\ee
For all $\alpha,\beta \in \mathbb{R}$, $\alpha^2+\beta^2 \neq 0$, this yields 
a real and regular solution of the KPI equation, representing a periodic chain of `lumps' 
(see Fig.~1 and \cite{periodic}). Also $\alpha \mapsto \imath \alpha$ (instead of 
$\beta \mapsto \imath \beta$) gives a real and regular KPI solution if $\alpha >0$. 
In the latter case we have $R = -L^\dagger$ and $K$ is Hermitian. For $M=N$ and KPI 
($t_{2n} \mapsto \imath \, t_{2n}$), these conditions, together with an invertible 
Hermitian $C$, guarantee that $\tau$ is \emph{real}. The rank condition can then 
be written as $R K - K L = w w^\dagger$. As a consequence, 
$B = \int_{-\infty}^{t_1} B_{t_1} dt_1 = -C \int_{-\infty}^{t_1} \tilde{w} \tilde{w}^\dagger dt_1$ 
with $\tilde{w} := e^{-\xi(R)} w$, provided that $B$ vanishes as $t_1 \to -\infty$. 
Then $\tau >0$ if the components of $\tilde{w}$ are linearly independent and if $C$ 
is negative definite (cf. proposition~4 in \cite{LM}). 
This ensures the absence of singularities of $u$. 
An example with non-diagonalizable $L$ is given by
\bea
   L = \left( \begin{array}{cc} \alpha + \imath \, \beta & 1 \\
               0 & \alpha + \imath \, \beta \end{array} \right) \, , \; 
   K = \left( \begin{array}{rr} -\frac{1}{2 \alpha} & \frac{1}{4 \alpha^2} \\
         \frac{1}{4 \alpha^2} & - \frac{1}{4 \alpha^3} \end{array} \right) \, , \; 
   C = - \left( \begin{array}{cc} \frac{16 \alpha^4}{\gamma} & 0 \\ 0 & \gamma \end{array} \right) \, , 
         \label{shark}
\eea
with $\alpha, \beta, \gamma \in \mathbb{R}$. Dropping a factor which does not influence $u$, 
we obtain
\bea
 \tau &=& \frac{1}{16} + \frac{4 \alpha^6}{\gamma^2} 
    + \frac{\alpha^2}{4} \left| t_1 - 2 \imath \, (\alpha - \imath \, \beta) \, t_2 
       + 3 \, (\alpha - \imath \, \beta)^2 \, t_3 \right|^2    \nonumber \\
 && + \frac{\alpha^3}{\gamma} \cosh\left( 2 \alpha \, (t_1 - 2 \beta \, t_2 
          + (\alpha^2 - 3 \beta^2) \, t_3) \right) \, , 
      \label{tau_sol+line}
\eea
which is positive if $\alpha,\gamma >0$. 
Some plots of corresponding KPI solutions are shown in Fig.~4. 
In the limit $\gamma \to \infty$, a KPI lump solution \cite{MZBIM} 
is recovered. Multi-lump generalizations \cite{MZBIM,LM,ACTV} can be obtained from 
$N \times N$ Jordan type matrices $L,R$. 
\bfg[t] 
\bc 
\resizebox{!}{4cm}{
\includegraphics{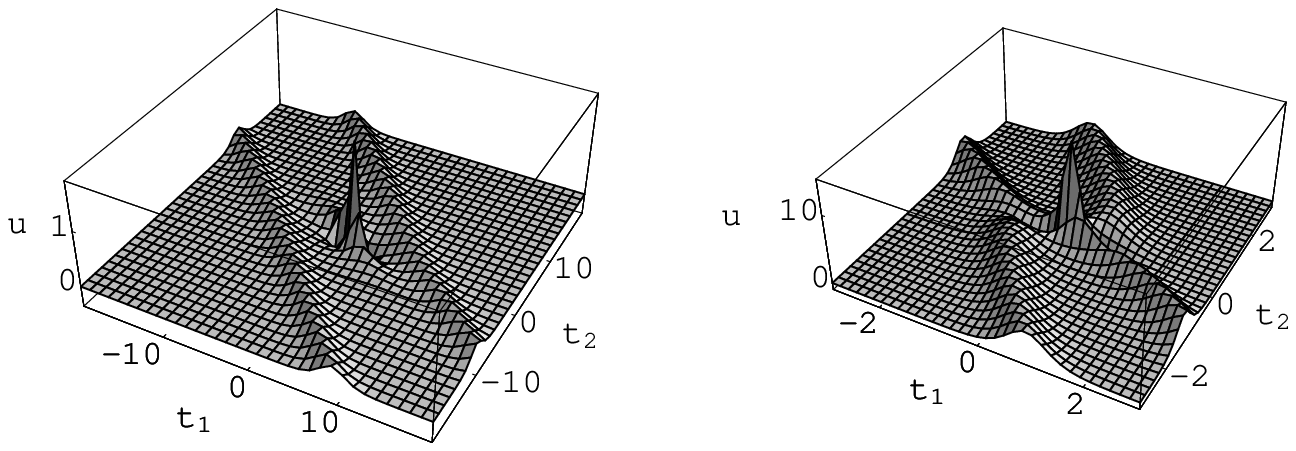}
}
\ec 
\vspace{-10mm}
\caption{Snapshots at $t_3=0$ of KPI solutions determined by (\ref{shark}). 
For $\alpha=\frac{4}{7}, \beta=-\frac{1}{2},\gamma=300$, a lump moves between two line solitons 
(left plot). 
For $\alpha=2, \beta=-\frac{1}{2},\gamma=100$, the right line soliton develops a 
lump, shortly after swallowed by the left one (right plot).}
\efg

\section{Final remarks}
So far we only looked at the special case where $\A'$ is a matrix algebra and 
$g =0$ (see the end of section~1 and the first of eqs (\ref{nu_augm})). 
If $g \neq 0$, the nonassociative hierarchy of ODEs leads to more complicated 
Riccati equations which, however, can also be solved. We plan to present a 
corresponding analysis elsewhere. Theorem~2 offers still further 
possibilities to obtain solutions of KP hierarchies.

\end{document}